\documentclass[11pt]{article}
\usepackage{epsfig,apjfonts,multicol}
\def\gtrsim{\mathrel{\hbox{\rlap{\hbox{\lower4pt\hbox{$\sim$}}}\hbox{$>$}}}}
\def\lesssim{\mathrel{\hbox{\rlap{\hbox{\lower4pt\hbox{$\sim$}}}\hbox{$<$}}}}
\begin{document}

\vspace*{-0.50in}
\hspace*{5.15in}
{\small\sl November 10, 2010}

\vspace*{-0.06in}
\begin{center}
{\Large\bf Ultraviolet Coronagraph Spectroscopy:$\,$
A Key Capability for Understanding

\vspace*{0.03in}
the Physics of Solar Wind Acceleration}

\vspace*{0.06in}
S.~R.~Cranmer,$^{1}$
J.~L.~Kohl,$^{1}$
D.~Alexander,$^{2}$
A.~Bhattacharjee,$^{3}$
B.~A.~Breech,$^{4}$
N.~S.~Brickhouse,$^{1}$
B.~D.~G.~Chandran,$^{3}$
A.~K.~Dupree,$^{1}$
R.~Esser,$^{5}$
S.~P.~Gary,$^{6}$
J.~V.~Hollweg,$^{3}$
P.~A.~Isenberg,$^{3}$
S.~W.~Kahler,$^{7}$
Y.-K.~Ko,$^{8}$
J.~M.~Laming,$^{8}$
E.~Landi,$^{9}$,
W.~H.~Matthaeus,$^{10}$
N.~A.~Murphy,$^{1}$
S.~Oughton,$^{11}$
J.~C.~Raymond,$^{1}$
D.~B.~Reisenfeld,$^{12}$
S.~T.~Suess,$^{13}$
A.~A.~van~Ballegooijen,$^{1}$
and
B.~E.~Wood$^{8}$

\vspace*{0.02in}
{\em
$^{1}$Harvard-Smithsonian CfA,
$\,$
$^{2}$Rice University,
$\,$
$^{3}$U.\  New Hampshire,
$\,$
$^{4}$U.S.\  Army Research Laboratory,
$\,$
$^{5}$U.\  Troms{\o}, Norway,
$\,$
$^{6}$Los Alamos National Laboratory,
$\,$
$^{7}$Air Force Research Laboratory,
$\,$
$^{8}$Naval Research Laboratory,
$\,$
$^{9}$U.\  Michigan,
$\,$
$^{10}$Bartol Research Institute, U.\  Delaware,
$\,$
$^{11}$U.\  Waikato, New Zealand,
$\,$
$^{12}$U.~Montana,
$\,$
$^{13}$National Space Science \& Technology Center
}
%-----------------------solicited----------------
% Miralles
% Hollweg       -- YES
% Chandran      -- YES
% Isenberg      -- YES
% Bhattacharjee -- YES
% Matthaeus     -- YES
% E Landi       -- YES
% Suess         -- YES
% Esser         -- YES
% Lie-Svendsen  -- NO
% Laming        -- YES
% Brickhouse    -- YES
% SP Gary       -- YES
% Moses
% Kahler        -- YES
% Nick Murphy   -- YES
% Fleck         -- NO
% Brian Wood    -- YES
% Jim Drake     -- NO?
% Justin Kasper -- NO
% D. Reisenfeld -- YES
% Paul Janzen
% Kuen Ko       -- YES
% D. Alexander  -- YES
% S. Spangler
% S. Oughton    -- YES
% B. Breech     -- YES
% A. Dupree     -- YES
% E. Avrett
% Randall Smith -- NO
%------------------not yet solicited-------------
% Dmitruk, others at Bartol
% George Doschek, Russ Howard
% Fineschi, Giordano, Romoli, Noci, Ciaravella, Poletto
% Huber
% Greg Howes

\vspace*{0.14in}
{\bf Abstract}
\end{center}

\vspace*{-0.10in}
{\noindent
Understanding the physical processes responsible for accelerating
the solar wind requires detailed measurements of the collisionless
plasma in the extended solar corona.
Some key clues about these processes have come from instruments that
combine the power of an ultraviolet (UV) spectrometer with an
occulted telescope.
This combination enables measurements of ion emission lines far
from the bright solar disk, where most of the solar wind acceleration
occurs.
Although the UVCS instrument on {\em SOHO} made several key discoveries,
many questions remain unanswered because its capabilities were limited.
This white paper summarizes these past achievements and also
describes what can be accomplished with next-generation instrumentation
of this kind.}

\vspace*{-0.03in}
\begin{center}
{\bf 1. Background and Motivation}
\end{center}

\vspace*{-0.08in}
The hot, ionized outer atmosphere of the Sun is a unique testbed
for the study of magnetohydrodynamics (MHD) and plasma physics,
with ranges of parameters that are inaccessible on Earth.
Although considerable progress has been made during the last few
decades, we still do not know the basic
physical processes responsible for heating the million-degree
solar corona and accelerating the solar wind.
Identifying these processes is important not only for
understanding the origins and impacts of space weather
(e.g., Schwenn 2006; Eastwood 2008),
but also for establishing a baseline of knowledge about a
well-resolved star that is directly relevant to other
astrophysical systems.

In order to construct and test theoretical models, a wide range of
measurements of relevant plasma parameters must be available.
In the low-density, open-field regions that reach into
interplanetary space, more parameters need to be measured
(in comparison to collision-dominated regions)
because the plasma becomes {\bf collisionless.}
In other words, individual particle species---e.g., protons,
electrons, helium, and minor ions---can exhibit different properties
from one another.
Such differences in particle velocity distributions are valuable
probes of ``microscopic'' processes of heating and acceleration.
In interplanetary space, such kinetic properties have been measured for
decades by {\em in~situ} particle instruments (e.g.,
Marsch 1999, 2006; Kasper et al.\  2008).
However, measurements in the near-Sun regions that are being actively
heated and accelerated have been more limited in scope.

Remote-sensing measurements of plasma properties in the so-called
``extended solar corona'' (above about 1.5 $R_{\odot}$ measured
from Sun-center) are
difficult to make because the photon emission at large heights is
fainter by many orders of magnitude than the emission from the solar
disk.
Standard telescopes, that do not explicitly block out the solar disk,
typically contain enough scattered light from the disk to
totally mask the faint off-limb emission.
Because the corona is highly ionized, the dominant spectral
features are at wavelengths accessible only from space.
For these reasons, a series of Ultraviolet Coronagraph Spectrometer
(UVCS) instruments have been built and flown on rockets, on a
Shuttle-deployed {\em Spartan} payload, and as an instrument on
the {\em Solar and Heliospheric Observatory} ({\em{SOHO}}) spacecraft;
see, e.g., Kohl et al.\  (1978, 1994, 1995, 1997, 2006).
In these instruments, the coronagraphic blocking of bright light
from the solar disk is done efficiently with a pair of
occulters: one external to the telescope and one internal.
Figure 1 illustrates UVCS/{\em{SOHO}} observations of the extended
corona.
\begin{figure}[!t]
\vspace*{0.01in}
\hspace*{0.05in}
\epsfig{figure=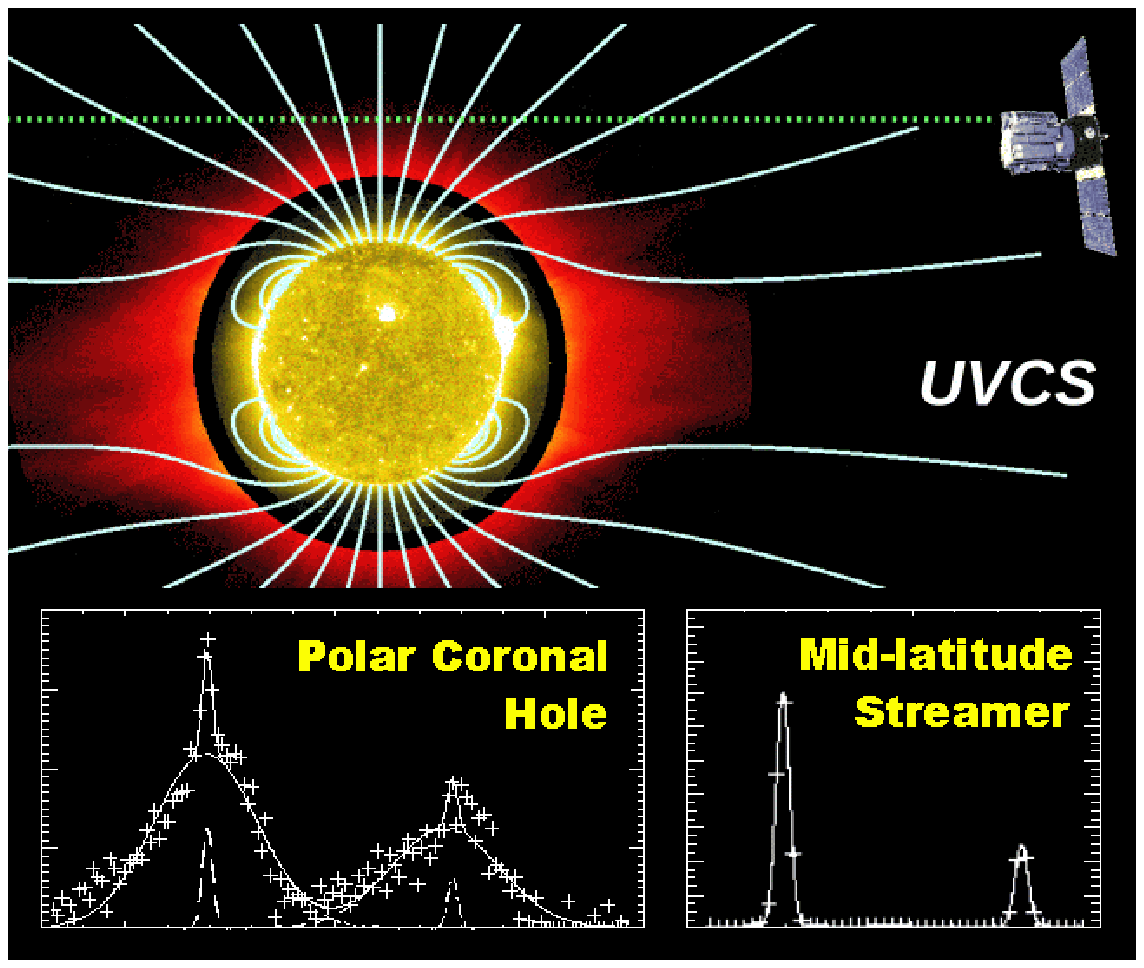,width=3.40in}

\vspace*{-2.43in}
\begin{tabular}{@{\hspace*{3.65in}}p{2.70in}}
\small
{\bf Figure 1:}
Combined {\em SOHO}
image of the solar corona from 17 August 1996, showing the
solar disk in Fe~XII 195~{\AA} intensity from EIT (yellow inner image)
and the extended corona in O~VI 1032~{\AA} intensity from UVCS (red
outer image).
A typical UVCS observational line of sight is shown in green.
Axisymmetric field lines are from the solar-minimum model of
Banaszkiewicz et al.\  (1998), and O~VI emission line profiles
(bottom) are from various UVCS observations at $r > 2 \, R_{\odot}$
in 1996--1997 (Kohl et al.\  1997).
\end{tabular}

\vspace*{0.45in}
\hspace*{0.10in}

\end{figure}

By measuring off-limb emission lines formed both by collisional excitation
and by the scattering of solar-disk photons, UV
spectroscopy provides a multi-faceted characterization of the
kinetic properties of atoms, ions, and free electrons (e.g.,
Withbroe et al.\  1982; Kohl et al.\  2006).
The Doppler-broadened shapes of emission lines are direct probes
of line-of-sight (LOS) particle velocity distributions (i.e.,
essentially providing $T_{\perp}$ when the off-limb magnetic field
is $\sim$radial).
Red/blue Doppler shifts reveal bulk flows along the LOS.
Integrated intensities of resonantly scattered lines can be
used to constrain the solar wind velocity and other details
about the velocity distribution in the radial direction
(e.g., $u_{\parallel}$ and $T_{\parallel}$); this is the so-called
``Doppler dimming/pumping'' diagnostic (e.g., Noci et al.\  1987).
Intensities of collisionally dominated lines---especially when
combined into an emission measure distribution---can
constrain electron temperatures, densities, and elemental
abundances in coronal plasma.
Even departures from Maxwellian and bi-Maxwellian velocity
distributions are detectable with spectroscopic measurements
having sufficient sensitivity and spectral resolution.

In the {\bf fast solar wind,} UVCS/{\em{SOHO}} measured outflow speeds
that become supersonic much closer to the Sun than previously believed.
In coronal holes, heavy ions (e.g., O$^{+5}$) were found to
flow faster than the protons, to be heated hundreds of times
more strongly than protons and electrons, and to have
anisotropic temperatures with $T_{\perp} > T_{\parallel}$
(Kohl et al.\  1997, 1998, 1999; Cranmer et al.\  1999b).
Specifically, the anisotropic and super-heated oxygen velocity
distributions (with $T_{\perp} > 10^{8}$ K) have guided theorists
to discard some candidate physical processes and further investigate
others.
Hollweg \& Isenberg (2002) stated in a review paper that
``We have seen that the information provided by UVCS has been
pivotal in defining how research has proceeded during the past
few years.''
\footnote{More recently, these results were highlighted
\rule{0in}{0.25in}%
in Chapter 3 of the 2010 report of the APS-sponsored
Workshop on Opportunities in Plasma Astrophysics (WOPA); see
http://www.pppl.gov/conferences/2010/WOPA/}

Figure 2 gives more information about how the temperatures of
protons, electrons, and an example minor ion species
(O$^{+5}$ in the corona, O$^{+6}$ {\em in~situ})
differ from one another in the high-speed solar wind.
Although the measured kinetic temperatures for protons are
of order 2--3.5 MK, these are combinations of random thermal motions
and ``nonthermal'' broadening due to Alfv\'{e}n waves
or other unresolved line-of-sight motions.
In Figure 2 we attempted to remove the contribution of waves
to show the true proton temperatures.
The resulting $T_p$ does not show as clear a signal of
preferential heating (relative to $T_e$) as one would have seen
from just the kinetic temperature.
Although one can still marginally see that $T_{p} > T_{e}$,
the existing measurements of $T_p$ and $T_e$ do not overlap
with one another in radius.
Improved measurements are needed in order to better constrain
the proton and electron heating rates in the corona.
The green points in Figure 2 indicate the key measurements of
preferential O$^{+5}$ heating in polar coronal holes.
The magenta and orange curves show a subset of the many theoretical
models that have been constructed to explain these data.
\begin{figure}[!t]
\vspace*{-0.02in}
\hspace*{-0.095in}
\epsfig{figure=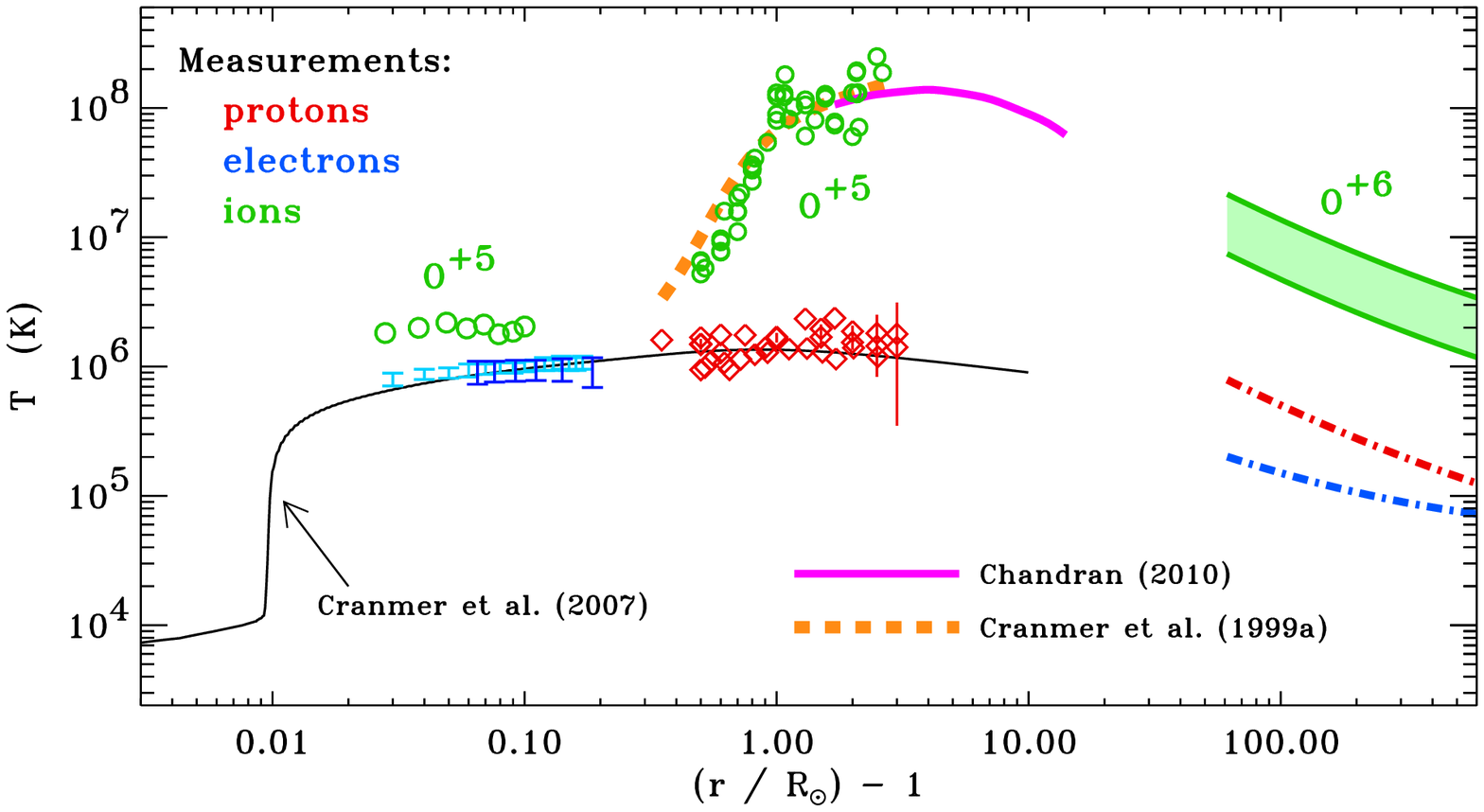,width=6.56in}

\small
{\bf Figure 2:}
Radial dependence of temperatures
in polar coronal holes and fast wind streams.
Mean plasma temperature from a turbulence-driven coronal heating
model (solid black curve; Cranmer et al.\  2007).
$T_e$ from off-limb SUMER measurements made by Wilhelm (2006)
(dark blue bars) and Landi (2008) (light blue bars).
$T_p$ from UVCS data assembled by Cranmer (2004) (red symbols).
Perpendicular O$^{+5}$ ion temperatures (green circles) from
Landi \& Cranmer (2009) ($r < 1.1 \, R_{\odot}$) and
Cranmer et al.\  (2008) ($r > 1.5 \, R_{\odot}$).
{\em In~situ} values of $T_p$ and $T_e$ (red and blue dot-dashed curves)
at $r > 60 \, R_{\odot}$ are from Cranmer et al.\  (2009), and the
{\em in~situ} range for O$^{+6}$ (light green region) is inferred
from, e.g., Collier et al.\  (1996).
Ion heating theories assuming high-$k_{\perp}$ waves (magenta solid curve)
and high-$k_{\parallel}$ waves (orange dashed curve) both appear to
succeed in modeling the O$^{+5}$ data.
\end{figure}

At solar minimum, UVCS/{\em{SOHO}} found that the {\bf slow solar wind}
flows mostly along the outer edges of bright streamers, near
locations with measured abundance patterns matching those of
the {\em in~situ} slow wind
(Raymond et al.\  1997; Strachan et al.\  2002; Abbo et al.\  2010).
The closed-field ``core'' regions of streamers, however, exhibit
heavy element abundances only 3\% to 30\% of those seen at 1 AU,
indicating gravitational settling (e.g., Raymond 1999;
V\'{a}squez \& Raymond 2005) or complex flux-tube geometries
(Noci et al.\  1997).
UVCS observed the transition from a high-density
collision-dominated plasma at low heights in streamers to a
low-density collisionless plasma at large heights, the latter
exhibiting high ion temperatures and anisotropies that suggest similar
physics as in the fast wind (Frazin et al.\  2003).
UVCS has also been used to measure plasma properties of
coronal mass ejections (CMEs) and put useful constraints on their
reconnection rates, 3D flow velocities, energy budgets, magnetic
helicities, and shock compression ratios (see, e.g., Raymond 2002;
Kohl et al.\  2006).
%The low-contrast ``blobs'' seen emerging from streamers in
%visible-light LASCO/{\em{SOHO}} movies appear to be a valuable tracer of
%slow wind acceleration, but the blobs themselves do not comprise
%a large fraction of the slow wind mass flux (e.g., Wang et al.\  2000).

%\vspace*{-0.01in}
\newpage

\vspace*{-0.36in}
\begin{center}
{\bf 2. Next-Generation Capabilities Are Needed}
\end{center}

\vspace*{-0.04in}
Despite the advances outlined above, the diagnostic capabilities
of UVCS were limited to what was foreseen before the {\em SOHO} era
(when only H~I Ly$\alpha$ had been observed).
Thus, we are still severely limited in our ability to answer the
fundamental questions by our lack of knowledge of the plasma
properties in the solar wind's acceleration region.
The following enhancements in measurement capability are needed to
be able to conclusively identify and characterize the
processes that energize the solar wind.

\vspace*{0.02in}
\noindent
\newcounter{bean}
\begin{list}{\arabic{bean}.}{\usecounter{bean}%
\setlength{\leftmargin}{0.16in}%
\setlength{\rightmargin}{0.0in}%
\setlength{\labelwidth}{0.16in}%
\setlength{\labelsep}{0.05in}%
\setlength{\listparindent}{0.0in}%
\setlength{\itemsep}{0.03in}%
\setlength{\parsep}{0.0in}%
\setlength{\topsep}{0.0in}}

\item[a.]
The key UVCS discovery of preferential ion heating and anisotropic
ion velocity distributions near the Sun was essentially limited to
just one ion (O$^{+5}$).
If the kinetic properties of {\bf additional minor ions} were to be
measured in the extended corona (i.e., a wider sampling of charge/mass
combinations) we could much better determine the nature of the
dominant collisionless heating process.
Specifically, these measurements would provide an improved empirical
description of kinetic waves and turbulence in the corona,
with which we could conclusively identify the type of fluctuations
(such as ion cyclotron waves, kinetic Alfv\'{e}n waves, or other
nonlinear turbulent modes) as well as their means of dissipation
(Hollweg \& Isenberg 2002; Cranmer 2002, 2009).
Sampling as broad a range as possible in the ion charge-to-mass ratio
($Z/A$) provides a sensitive probe of the heating; e.g.,
S$^{+5}$ (933~{\AA}, $Z/A = 0.16$).
Ca$^{+9}$ (557~{\AA}, $Z/A = 0.22$),
Si$^{+8}$ (296~{\AA}, $Z/A = 0.28$),
O$^{+5}$ (1032~{\AA}, $Z/A = 0.31$),
Mg$^{+9}$ (610~{\AA}, $Z/A = 0.37$),
and
Si$^{+11}$ (499~{\AA}, $Z/A = 0.39$).
With these ions, the two models shown in Figure 2
(which use either high-$k_{\perp}$ or high-$k_{\parallel}$ waves)
would be clearly distinguished from one another; see also Leamon
et al.\  (2000).
Our confidence in the uniqueness of a successful model increases as
the number of ions observed increases.

\item[b.]
UVCS provided new constraints on the heating of minor ions, but not
so much on the heating of the primary proton--electron plasma.
Observationally, the biggest missing piece is the
{\bf electron temperature} above $\sim$1.5 $R_{\odot}$.
Direct measurements of $T_e$ from, e.g., the broad Thomson-scattered
component of H~I Ly$\alpha$, would allow us to determine
the bulk-plasma heating rate in different solar wind regions,
as well as the partitioning of energy between protons and electrons.
This partitioning is a key diagnostic of turbulence models
(e.g., Matthaeus et al.\  2003; Howes 2010)
as well as a driver of the stability of
helmet streamers (e.g., Endeve et al.\  2004).
The downward conduction of electron thermal energy---from the corona
to the transition region---sets the base pressure and
thus the mass flux of the solar wind (Withbroe 1988;
Lie-Svendsen et al.\  2002), thus making
measurements of $T_{e}(r)$ especially important.
Improved observations of proton temperatures are also important,
since the measured rates of proton heating can be compared directly
with the energy available to protons in ion cyclotron waves (as constrained
by future measurements of additional minor ions; see item [a] above)
to reveal the relative importance of resonant wave heating to the
primary plasma.

\item[c.]
In addition to protons and electrons, {\bf helium} also plays a major
role in the ``primary'' plasma of the solar wind.
Helium may regulate the wind's mass flux (Hansteen et al.\  1994)
and set a lower limit on its outflow speed (Kasper et al.\  2007).
Preferential heating of alpha particles may even
dominate the coronal heating close to the Sun and control how much
heat is received by the protons (Liewer et al.\  2001; Li 2003).
Observations of the He~II 304~{\AA} and He~I 584~{\AA} lines can
be used to measure
the coronal helium abundance, departures from ionization
equilibrium, and the amount of preferential alpha-particle heating.

\item[d.]
Space-based observations by {\em Hinode} and {\em SDO}, as well
as ground-based eclipse observations, have highlighted the importance
of {\bf greater spatial and temporal resolution}
to the direct identification of MHD waves with periods between
about 0.1 and 10 minutes
(De Pontieu et al.\  2007; Pasachoff et al.\  2002;
Singh et al.\  2009).
The predicted amplitudes of most MHD waves grow significantly as
they propagate up to the extended corona, thus
making a UVCS-type instrument with greater sensitivity and resolution
ideal for detecting both compressive waves (via intensity oscillations) and
Alfv\'{e}n waves (via Doppler shifts).
Measuring the frequency and wavenumber dependence of these waves
would open new windows on our understanding of coronal turbulence
(e.g., Matthaeus et al.\  1999).
Greater spatial and temporal resolution also allows improved
characterization of small-scale inhomogeneities between neighboring flux
tubes in coronal holes and streamers.
Precise measurements of cross-field density gradients can put
firm constraints on models of coronal heating via drift instabilities
(e.g., Mecheri \& Marsch 2008) and MHD discontinuities in the solar
wind (Feldman et al.\  1997; Borovsky 2008).

\item[e.]
Recent analysis of {\em in~situ} data shows that the
combined {\bf elemental abundances and charge states} of solar wind
ions can be used to effectively probe the origin of heliospheric
wind streams in closed or open flux tubes (Zurbuchen 2007).
There have been some isolated comparisons of ion composition
between UVCS and {\em in~situ} measurements (e.g., Ko et al.\  2006),
but higher resolution and greater sensitivity to more ions is needed
to produce corona-heliosphere ``mappings'' of abundances and charge states
that are detailed enough to test models of coronal gravitational
settling and ion drag (Lenz 2004; Li et al.\  2006)
and the first ionization potential (FIP) effect (e.g., Laming 2004, 2009).
It is also important to measure a broader range of low-FIP and high-FIP
ions in order to better map out the way the solar atmosphere becomes
fractionated.

\item[f.]
Measuring {\bf non-Maxwellian velocity distributions} of
electrons and positive ions would provide even more stringent tests of
specific models of MHD turbulence, cyclotron resonance, and
velocity filtration.
An instrument with greater sensitivity than UVCS/{\em{SOHO}} could
detect subtle departures from Gaussian line shapes
that signal the presence of specific non-Maxwellian distributions
(e.g., Cranmer 1998, 2001).

\end{list}

\vspace*{0.04in}
\noindent
New capabilities such as these would be enabled by greater photon
sensitivity, an expanded wavelength range, and the use of measurements
that heretofore have been utilized only in a testing capacity
(e.g., Thomson-scattered H~I Ly$\alpha$ to obtain $T_e$).
These would allow
the relative contributions of different physical processes
to the heating and acceleration of all solar wind plasma
components to be determined directly.

\vspace*{0.02in}
\begin{center}
{\bf 3. Synergy with Other Instrumentation and Missions}
\end{center}

\vspace*{-0.02in}
UV coronagraph spectroscopy has been a key asset to a large number
of multi-instrument and multi-mission observation campaigns
(e.g., Galvin \& Kohl 1999).
For example,
coordinated observations taken at times when {\em SOHO} and
{\em Ulysses} could track the same plasma parcels from the Sun to the
outer heliosphere led to new methods of mapping field lines
over long distances in the solar wind (Suess et al.\  2000;
Poletto et al.\  2002).
Next-generation UVCS-type measurements would also provide key
context for the regions of the inner heliosphere that
{\em Solar Probe Plus} ({\em{SPP}}) and {\em Solar Orbiter} will
fly through,
and they would provide direct measurements of coronal regions inaccessible
even to {\em SPP} (i.e., $r < 9 \, R_{\odot}$).

Measurements from other instruments also can enhance the ability of
UV coronagraph spectroscopy to determine coronal plasma parameters.
Although UV emission lines can be used to measure the coronal
electron density (e.g., Noci et al.\  1997; Antonucci et al.\  2004),
the most straightforward method to obtain $n_{e}$ has been the analysis of
polarized brightness measured by a white-light coronagraph.
The combination of a UVCS-type instrument and a white-light imaging
coronagraph provides useful diagnostics that extend beyond what is
possible with either instrument in isolation.

There are several possible mission concepts for next-generation
UV coronagraph spectrometers.
Sub-orbital rockets are useful for demonstrating
the capabilities of this kind of instrument, as well as being able to
obtain useful scientific results (see {\S}~4.1 of Kohl et al.\  2006).
Studies have also shown that a standard Low-Earth orbit
(350--600 km altitude at 28$^{\circ}$ latitude) can provide adequate
observation time, low UV optical depth from the Earth's atmosphere,
and negligible contamination
to enable the full range of required UV coronagraph spectroscopy.
A polar, Sun-synchronous orbit would of course be an improvement over
a low-latitude orbit because of the lack of diurnal solar occultation by
the Earth.
Another possibility is to position the spacecraft at the
{\bf Earth-Sun L5 point} (see also Vourlidas et al.\  2010)
which eliminates issues of atmospheric absorption and orbital night.
This location also positions the prime off-limb field of view
(i.e., the ``plane of the sky'') close to the Earth-Sun line,
so that geoeffective solar wind streams and CMEs can be measured directly.

\newpage

\vspace*{-0.36in}
%\vspace*{0.02in}
\begin{center}
{\bf 4. Instrumentation and Costs}
\end{center}

\vspace*{-0.02in}
The UV spectroscopic observations required to produce
detailed empirical descriptions of solar wind source regions can be
accomplished with a large-aperture ultraviolet coronagraph spectrometer
(LAUVCS).
The large aperture is needed to provide the sensitivity to observe
relatively weak spectral lines with the required range of
charge-to-mass ratios.
The large aperture can be realized with a remote external occulter
supported by a deployable mast (see Figure 3).
Such an instrument would accommodate a much larger 
telescope area than UVCS/{\em{SOHO}} that is both shielded from direct 
solar disk rays and also has an unobstructed view of the coronal 
heights of interest.
When equipped with state-of-the-art reflection coatings, high
ruling efficiency diffraction gratings, and array detectors, such an
instrument exceeds the sensitivity and stray light suppression
needed to make all required coronal observations
in both streamers and coronal holes up to
at least 5 $R_{\odot}$ and often higher.
It is also possible to achieve the large aperture 
with an internally occulted coronagraph where the solar-disk radiation 
impinges directly on the telescope primary.
The sensitivity is just as good as that of the externally occulted
instrument, but the stray light characteristics only allow
observations up to $\sim 2 \, R_{\odot}$ in coronal holes and
$\sim 3 \, R_{\odot}$ in streamers.
Compared to UVCS, either instrument would make a tremendous step
forward in describing solar wind source regions and identifying
the physical processes that heat and accelerate the solar wind
(and CMEs\footnote{The ability of next-generation UV coronagraph
\rule{0in}{0.25in}%
spectroscopy to answer fundamental questions about how CMEs and
solar energetic particles (SEPs) are
produced will be described in a separate white paper.}).
\begin{figure}[!t]
\hspace*{0.033in}
\epsfig{figure=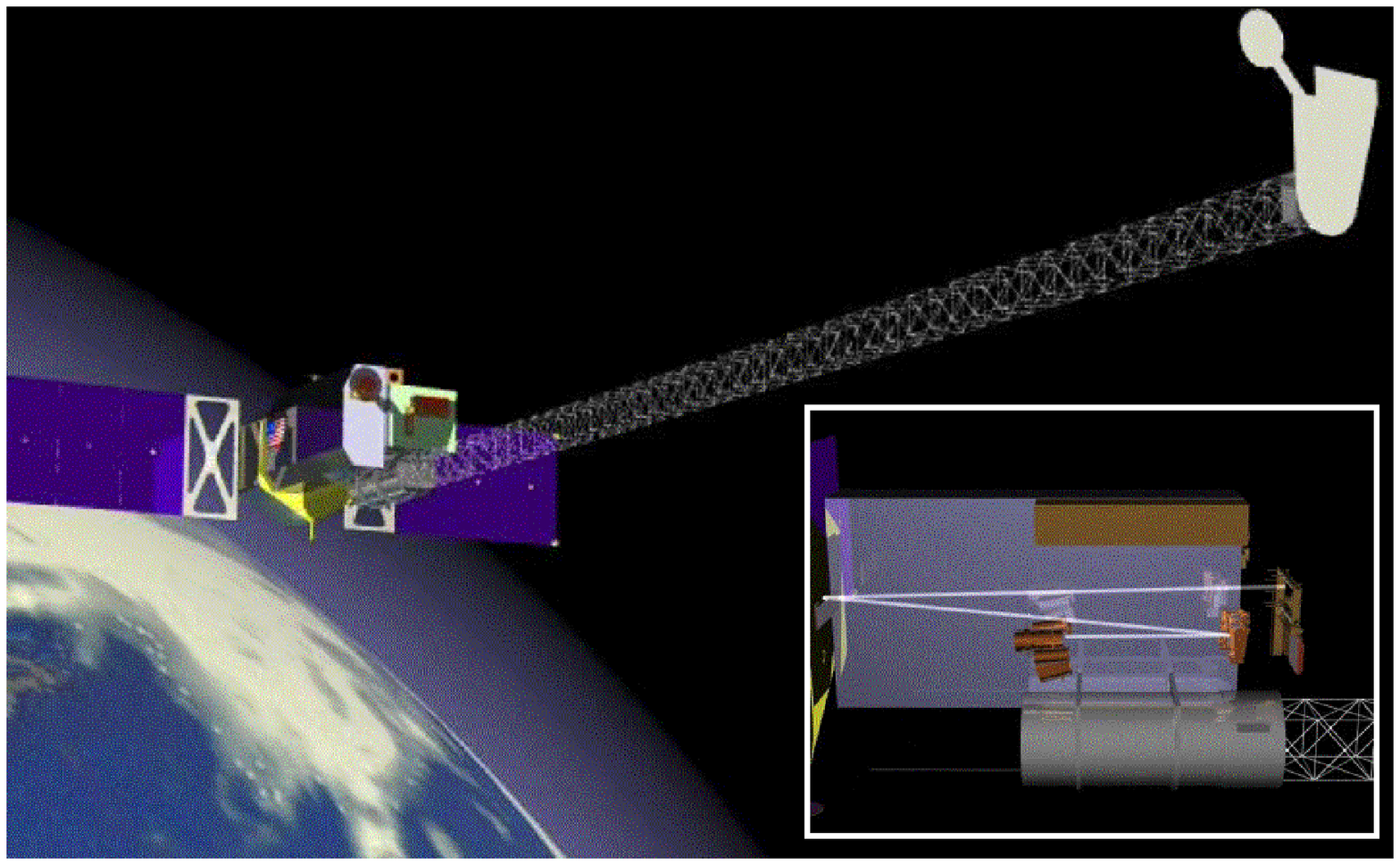,width=6.33in}

\vspace*{0.03in}
\small
{\bf Figure 3:}
Spacecraft concept for a large-aperture UV coronagraph
with an external occulter at the end of a 13~m boom.
The inset shows a diagram of an advanced LAUVCS instrument
(see Kohl et al.\  2006).
\end{figure}

Both externally occulted and internally occulted LAUVCS type instruments
can be built with existing technology.
Deployable masts to support the remote external occulter, such as
the one baselined for the {\em NuSTAR} mission, are commercially 
available.
There is also ``in~space'' performance data that demonstrates
the ATK ADAM mast exceeds the specifications required for the
externally occulted LAUVCS.
Intensified CCD (ICCD) detectors similar to those used on
{\em XMM} and {\em Swift} can be used for a LAUVCS, and 
they are capable of meeting all requirements for the solar wind science 
while maintaining the full instrument sensitivity.
The KBr-coated microchannel plates
that are needed to meet the UV sensitivity requirements are 
also available commercially, and they can be incorporated into an
ICCD with an acceptable air exposure.
To achieve the LAUVCS science objectives for CMEs---without the need
to attenuate the associated high light levels---it is highly desirable to
use an intensified active pixel sensor (IAPS), which is the next step
beyond the ICCD.
A laboratory model of such a device exists at the Mullard Space 
Science Laboratory (MSSL), and it has been clocked at 66~MHz to achieve the 
desired maximum detection rate of 32~Hz per pixel with less than a 10\% 
dead time loss.

Both the solar wind and the baseline CME science objectives can be
achieved with an ICCD detector similar to those flown on {\em XMM}
and {\em Swift.}
The advanced technology IAPS detectors currently being developed at
MSSL are showing great promise, and they could provide a dynamic range
and maximum count rate capability that would accommodate the
bright CME emissions without the need for attenuators.

The cost (in FY 2011 dollars) of an externally occulted LAUVCS
for a Class~C mission
with limited redundancy, including design, development,
calibration, environmental testing, and support for integration, is
approximately \$45M including the cost of the deployable mast.
The cost of the internally occulted instrument would be about 
\$15M less.
These estimated costs do not include any reserve/contingency
or margin except for costed schedule reserve, and they do not include
any Phase~E mission operation and data analysis costs.

\vspace*{-0.15in}
\begin{center}
{\bf References}
\end{center}

\vspace*{-0.18in}
\footnotesize
\parindent=0.0in
\baselineskip=8.86pt
\setlength{\columnsep}{0.50cm}
\begin{multicols}{2}

Abbo, L., et al. 2010, Adv.\  Sp.\  Res., 46, 1400

Antonucci, E., et al. 2004, A\&A, 416, 749

%Axford, W. I., et al. 1999, Space Sci.\  Rev., 87, 25

Banaszkiewicz, M., et al. A\&A, 337, 940

Borovsky, J. E. 2008, JGR, 113, A08110

Chandran, B. D. G. 2010, ApJ, 720, 548

Collier, M. R., et al. 1996, GRL, 23, 1191

Cranmer, S. R. 1998, ApJ, 508, 925

Cranmer, S. R. 2001, JGR, 106, 24937

Cranmer, S. R. 2002, Proc.\  SOHO-11, 361, astro-ph/0209301

Cranmer, S. R. 2004, Proc.\  SOHO-15, 154, astro-ph/0409724

Cranmer, S. R. 2009, Living Rev.\  Sol.\  Phys., 6, 3

Cranmer, S. R., et al. 1999a, ApJ, 518, 937

Cranmer, S. R., et al. 1999b, ApJ, 511, 481

Cranmer, S. R., et al. 2007, ApJS, 171, 520

Cranmer, S. R., et al. 2008, ApJ, 678, 1480

Cranmer, S. R., et al. 2009, ApJ, 702, 1604

%Cranmer, S. R., \&  van Ballegooijen, A. A. 2010, ApJ, 720, 824

De Pontieu, B., et al. 2007, Science, 318, 1574

Eastwood, J. P. 2008, Phil.\  Trans.\  Roy.\  Soc.\  A, 366, 4489

Endeve, E., et al. 2004, ApJ, 603, 307

Feldman, W. C., et al. 1997, JGR, 102, 26905

Frazin, R. A., et al. 2003, ApJ, 597, 1145

Galvin, A. B., \& Kohl, J. L. 1999, JGR, 104, 9673

%Greco, A., et al. 2009, Phys.\  Rev.\  E, 80, 046401

Hansteen, V. H., et al. 1994, ApJ, 428, 843

Hollweg, J. V., \& Isenberg, P. A. 2002, JGR, 107 (A7), 1147

Howes, G. G. 2010, MNRAS, in press, arXiv:1009.4212

Kasper, J. C., et al. 2007, ApJ, 660, 901

Kasper, J. C., et al. 2008, PRL, 101, 261103

%Ko, Y.-K., et al. 2002, ApJ, 578, 979

%Ko, Y.-K., et al. 2005, ApJ, 623, 519

Ko, Y.-K., et al. 2006, ApJ, 646, 1275

Kohl, J. L., et al. 1978, in New Instrumentation for Space \\
\hspace*{0.16in} Astronomy (Oxford: Pergamon), 91

Kohl, J. L., et al. 1994, Space Sci.\  Rev., 70, 253

Kohl, J. L., et al. 1995, Solar Phys., 162, 313

Kohl, J. L., et al. 1997, Solar Phys., 175, 613

Kohl, J. L., et al. 1998, ApJ, 501, L127

Kohl, J. L., et al. 1999, ApJ, 510, L59

Kohl, J. L., et al. 2006, A\&A Review, 13, 31

Laming, J. M. 2004, ApJ, 614, 1063

Laming, J. M. 2009, ApJ, 695, 954

Laming, J. M., \& Lepri, S. T. 2007, ApJ, 660, 1642

Landi, E. 2008, ApJ, 685, 1270

Landi, E., \& Cranmer, S. R. 2009, ApJ, 691, 794

Leamon, R. J., et al. 2000, ApJ, 537, 1054

Lenz, D. D. 2004, ApJ, 604, 433

Li, B., et al. 2006, JGR, 111, A08106

Li, X. 2003, A\&A, 406, 345

Lie-Svendsen, {\O}., et al. 2002, ApJ, 566, 562

Liewer, P., et al. 2001, JGR, 106, 29261

Marsch, E. 1999, Space Sci.\  Rev., 87, 1

Marsch, E. 2006, Liv.\  Rev.\  Sol.\  Phys., 3, 1

Matthaeus, W. H., et al.\  1999, ApJ, 523, L93

Matthaeus, W. H., et al.\  2003, Nonlin.\  Proc.\  Geophys., 10, 93

Mecheri, R., \& Marsch, E. 2008, A\&A, 481, 853

Noci, G., et al. 1987, ApJ, 315, 706

Noci, G., et al. 1997, Proc.\  SOHO-5, ESA SP-404, 75

Pasachoff, J. M., et al. 2002, Solar Phys., 207, 241

Poletto, G., et al. 2002, JGR, 107, 1300

Raymond, J. C. 1999, Space Sci.\  Rev., 87, 55

Raymond, J. C. 2002, Proc.\  SOHO-11, ESA SP-508, 421

Raymond, J. C., et al. 1997, Solar Phys., 175, 645

Schwenn, R. 2006, Liv.\  Rev.\  Sol.\  Phys., 3, 2

Singh, J., et al. 2009, Solar Phys., 260, 125

%Strachan, L., et al. 1999, Space Sci.\  Rev., 87, 311

Strachan, L., et al. 2002, ApJ, 571, 1008

Suess, S. T., et al. 2002, JGR, 105, 25033

%Tu, C.-Y., \& Marsch, E. 1997, Solar Phys., 171, 363

V\'{a}squez, A. M., \& Raymond, J. C. 2005, ApJ, 619, 1132

Vourlidas, A., et al. 2010, NRC Decadal Survey White Paper

%Wang, Y.-M., et al. 2000, JGR, 105, 25133

Wilhelm, K. 2006, A\&A, 455, 697

Withbroe, G. L. 1988, ApJ, 325, 442

Withbroe, G. L., et al. 1982, Space Sci.\  Rev., 33, 17

Zurbuchen, T. H. 2007, Ann.\  Rev.\  Astron.\  Astrophys., 45, 297
\end{multicols}

\end{document}